\documentclass[12pt]{article}  
\usepackage{amssymb,eufrak}
\newcommand{\nc}{\newcommand}
\nc{\la}{\lambda} \nc{\alf}{\alpha}
\nc{\tht}{\theta}  \nc{\be}{\beta}  \nc{\eps}{\epsilon}
\nc{\ga}{\gamma}  \nc{\Ga}{\Gamma}  \nc{\De}{\Delta}
\nc{\de}{\delta} \nc{\si}{\sigma}  \nc{\ka}{\kappa}
\nc{\om}{\omega}  \nc{\qq}{\quad\quad}
\nc{\nf}{\infty}   \nc{\dl}{\mathop{\smash{\cal L}}}
\nc{\ra}{\rightarrow}  \nc{\ol}{\overline}
\nc{\beq}{\begin{equation}}  \nc{\barr}{\begin{array}}
\nc{\earr}{\end{array}}
\nc{\eeq}{\end{equation}}
\nc{\beqa}{\begin{eqnarray}}  \nc{\dst}{\displaystyle}
\nc{\eeqa}{\end{eqnarray}} \nc{\nnb}{\nonumber}
\nc{\bs}{\backslash}        \nc{\mbb}{\mathbb}
\nc{\brm}{\begin{remunerate}} \nc{\erm}{\end{remunerate}}
\nc{\nn}{\nonumber \\}
\nc{\p}[1]{{(\ref{#1})}}
\newcounter{muni}
{\usecounter{muni}
\setlength{\leftmargin}{0pt}\setlength{\itemindent}{38pt}

\newenvironment{remunerate}{\begin{list}{{\rm \arabic{muni}.}}
{\usecounter{muni}
\setlength{\leftmargin}{0pt}\setlength{\itemindent}{38pt}}}{\end{list}}
\topmargin=-2cm\textheight=23.5cm\textwidth=16.cm
\oddsidemargin=0.25cm\evensidemargin=0.25cm
\begin{document}
\begin{titlepage}
\begin{flushright}
LPTHE 01-16 \\
JINR E2-2001-70 \\
hep-th/0104078 \\
April 2001
\end{flushright}
\vskip 1.0truecm
\begin{center}
{\large \bf Quaternionic Extension of the Double Taub-NUT Metric}
\end{center}
\vskip 1.0truecm
\centerline{\bf Pierre-Yves Casteill${}^{\;a,1}$, Evgeny
Ivanov${}^{\;b,a, 2}$, Galliano Valent${}^{\;a,3}$} 
\vskip 1.0truecm
\centerline{${}^{a}$ \it
Laboratoire de Physique Th\'eorique et des Hautes Energies,}
\centerline{\it Unit\'e associ\'ee au CNRS URA 280,~Universit\'e Paris 7}
\centerline{\it 2 Place Jussieu, 75251 Paris Cedex 05, France}
\vskip5mm
\centerline{${}^{b}$\it Bogoliubov Laboratory of Theoretical
Physics, JINR,}
\centerline{\it Dubna, 141 980 Moscow region, Russia}

\vskip 1.0truecm  \nopagebreak

\begin{abstract}
\noindent Starting from the generic harmonic superspace action of the
quaternion-K\"ahler sigma models and using the quotient 
approach we present, in an explicit form, a  quaternion-K\"ahler extension of
the  double Taub-NUT metric. It possesses $U(1)\times U(1)$
isometry and supplies a new example of non-homogeneous Einstein metric with
self-dual Weyl tensor.
\end{abstract}
\vfill
{\it E-Mail:}\\
{\it 1) casteill@lpthe.jussieu.fr}\\
{\it 2) eivanov@thsun1.jinr.ru, eivanov@lpthe.jussieu.fr}\\
{\it 3) valent@lpthe.jussieu.fr}
\end{titlepage}

\noindent{\bf 1. Introduction}. In view of the distinguished role of
hyper-K\"ahler (HK) and quaternion-K\"ahler (QK) manifolds in string theory
(see, e.g., \cite{1}-\cite{2}), it is important to know the explicit form of
the corresponding  metrics. One of the approaches to this problem proceeds
from the generic actions of bosonic nonlinear sigma models with the HK or QK
targets.

A generic action for the bosonic QK sigma models was 
constructed in \cite{IvV}, based upon the well-known one-to-one correspondence
\cite{bw} between the QK manifolds and   local $N=2, d=4$ supersymmetry. 
This relationship was made manifest in \cite{gios3,gio}, where the most
general off-shell action for the hypermultiplet $N=2$ sigma models coupled to
$N=2$ supergravity was constructed in the framework of
$N=2$ harmonic superspace (HSS) \cite{hss}. The  generic QK sigma model
bosonic action was derived in \cite{IvV} by discarding the fermionic fields
and part of the bosonic ones in the general HSS sigma model action. The action
of physical bosons parametrizing the target QK manifold arises, like in the HK
case \cite{gios2}, after elimination of infinite sets of auxilairy fields
present in the off-shell hypermultiplet superfields. This amounts to solving
some differential equations on the internal sphere $S^2$ of the $SU(2)$
harmonic variables. It is a difficult problem in general to solve such
equations. As was shown in  \cite{IvV}, in the case of metrics with isometries
the computations can be greatly simplified  by using the HSS version of the QK
quotient construction \cite{orig,gal1}. An attractive feature of the HSS
quotient is that the isometries of the corresponding metric come out 
as manifest internal symmetries of the HSS sigma model action.  

In \cite{IvV}, using these techniques, we explicitly constructed QK extensions
of the Taub-NUT and Eguchi-Hanson (EH) HK metrics \cite{Rep}.
In this note we apply the HSS quotient approach  to construct a   QK extension
of  the 4-dimensional ``double Taub-NUT'' HK metric. The latter was derived
from the HSS approach in \cite{gorv} by directly solving the corresponding
harmonic differential  equations. It turns out that the HSS quotient allows
one to reproduce the same answer much easier, and it remarkably works in the
QK case as well. We gauge two commuting $U(1)$ symmetries of the system of
three ``free'' hypermultiplets and, after solving two algebraic constraints
and fully fixing gauges, are left with a 4-dimensional QK metric having two 
$U(1)$ isometries and going onto the double Taub-NUT in the HK limit. It is a
new explicit example of non-homogeneous QK metrics. Based on the results of
Przanowski \cite{Pr}, Tod \cite{To}  and Flaherty \cite{Fl}, this metric gives
also a new explicit solution of the coupled Einstein-Maxwell system with
self-dual Weyl tensor. 
\vspace{0.3cm}

\noindent{\bf 2. The gauged HSS action of the QK double Taub-NUT}. Details of
the general construction can be found in \cite{IvV}. Here we  apply the HSS
quotient approach to explicitly construct a sigma model  giving rise to a QK
generalization of the ``double Taub-NUT'' HK metric.  The latter belongs to
the class of two-center ALF metrics with the $U(1)\times U(1)$ isometry (one
$U(1)$ is triholomorphic)  and was treated in the HSS approach in \cite{gorv}.
     
We start with the action of three  hypermultiplet superfields,
\beq
Q^{+ a}_A (\zeta), \;\;g^{+ r}(\zeta)~, \qquad a=1,2; \; r = 1,2; \;A= 1,2,
\label{triad} \eeq 
possessing no any self-interaction. So, by reasoning of
\cite{bgio,IvV}, this action corresponds to the ``flat'' QK manifold ${\mbb
H}H^{3} \sim Sp(1,3)/Sp(1)\times Sp(3)$. In \p{triad}, the indices $a$ and $r$
are the doublet indices of two Pauli-G\"ursey-type
$SU(2)\,$s realized on $Q^{+a}_A$ and $g^{+r}$, the index $A$  is an extra
$SO(2)$ index. These superfields are given on the harmonic analytic $N=2$
superspace \beq
(\,\zeta \,) = (\,x^m, \theta^{+\mu}, \bar\theta^{+\dot\mu}, u^{+ i}, u^{-
k}\,)~, \label{anal}  
\eeq 
the coordinates $u^{+ i}, u^{- k}, \;u^{+i}u^-_i = 1,\; i,k = 1,2,$ being the
$SU(2)/U(1)$ harmonic variables, and they satisfy the pseudo-reality
conditions   \beq
(a)\; Q^{+ a}_A \equiv \widetilde{(Q^+_{a\,A})} = \epsilon^{ab}Q^{+}_{b\,A}~,
\quad  (b) \;g^{+ r} \equiv \widetilde{(g^+_r)} = \epsilon^{rs}g^{+}_s~,
\label{real} \eeq
where $\epsilon^{ab}\epsilon_{bc} = \delta^a_c\,, \;\epsilon^{12} = -1$. 
The generalized conjugation $\;\widetilde{}\;$ is the product of the ordinary
complex conjugation and a Weyl reflection of the sphere $S^2 \sim SU(2)/U(1)$
parametrized by $u^{\pm i}$. In the QK sigma model action below we shall need
only the bosonic components  in the $\theta$-expansion of the above
superfields:     
\beqa 
Q^{+ a}_A(\zeta) &=&
F^{+ a}_A(x,u) + i(\theta^+\sigma^m\bar\theta^+)B^{- a}_{m\,A}(x,u) +
(\theta^+)^2(\bar\theta^+)^2 G^{(-3 a)}_A(x,u)~, \nn 
g^{+ r}(\zeta) &=& g_0^{+ r}(x,u) + i(\theta^+\sigma^m\bar\theta^+)g^{-
\,r}_{m}(x,u) + (\theta^+)^2(\bar\theta^+)^2 g^{(-3 \,r)}(x,u) \label{thetaexp1} 
\eeqa
(possible terms $\sim (\theta^+)^2$ and $\sim (\bar\theta^+)^2$ can be shown
not to contribute to the final action). The component fields
still have a general harmonic expansion off shell. The physical bosonic
components $F^{ai}_A(x), g^{ri}(x)$ are defined as the first components in
the harmonic expansions of $F^{+ a}_A(x, u)$ and $g^{+r}_0(x, u)$    \beqa &&
F^{+a}_A(x,u) = F^{ai}_A(x)u^+_i + \cdots~, \quad  g_0^{+r}(x,u) =
g^{ri}(x)u^+_i + \cdots~, \nn   &&\overline{(F^{ai}_A(x))} =
\epsilon_{ab}\epsilon_{ik}F^{bk}_A(x)~, \; \overline{(g^{ri}(x))} =
\epsilon_{rs}\epsilon_{ik}g^{sk}(x)~.  
\eeqa	

The selection of two commuting $U(1)$ symmetries  to be
gauged and the form of the final gauge-invariant HSS action are uniquely 
determined by the natural requirement that the resulting action have two 
different limits corresponding to the earlier considered HSS quotient actions
of  the QK extensions of Taub-NUT and EH metrics \cite{IvV}. The
full action $S_{dTN}$ has the following
form   
\beqa  
S_{dTN} &=& {1\over 2}\int d\zeta^{(-4)}{\cal L}^{+4}_{dTN} 
- {1\over 2\kappa^2}\int d^4x \left[D(x) + V^{m(ij)}(x)V_{m(ij)}(x)\right]~,
\label{dTN} \\
{\cal L}^{+4}_{dTN} &=& -q^+_a{\cal
D}^{++}q^{+a} +  \kappa^2 (u^-\cdot q^+)^2\left[  Q^{+}_{aA}{\cal
D}^{++}Q^{+a}_{A} + g^+_r{\cal D}^{++}g^{+r} \right.\nn 
&& \left.  + \, W^{++}\left( Q^{+a}_AQ^+_{a B}\epsilon_{AB} - \kappa^2
c^{(ij)}g^+_ig^+_j + c^{(ij)}v^{+}_iv^+_j \right) \right. \nn
&& \left.+ \,V^{++}\left( 2(v^+\cdot g^+) -
a^{(ab)}Q^+_{aA}Q^+_{bA}\right) \right]~. \label{action2}  
\eeqa
Here, $d \zeta^{(-4)} = d^4x d^2\theta^+ d^2\bar\theta^+ du$ is the
measure of integration over \p{anal}, $(a\cdot b) \equiv a_i b^i$, the
covariant harmonic derivative ${\cal D}^{++}$ is defined by 
\beq
{\cal D}^{++} = D^{++} +
(\theta^+)^2(\bar\theta^+)^2\,
\{\; D(x)\,\partial^{--} + 6\,V^{m\,(ij)}(x)u^-_iu^-_j \,\partial_m \; \}~,
\label{substit}
\eeq
with $D^{++} = \partial^{++} - 2i\theta^+\sigma^m\bar\theta^+\,\partial_m$, 
$\partial^{\pm\pm} = u^{\pm i}/\partial u^{\mp i}$, the non-propagating
fields $D, V^{m(ij)}$ are inherited from  the $N=2$ supergravity Weyl
multiplet, $\kappa^2$ is the Einstein constant (or, from  the geometric
point of view, the parameter of contraction to  the HK case) and the new
harmonic $v^{+ i}$ is defined by   $$
v^{+ a} = \frac{q^{+a}}{(u^- \cdot q^{+})} = u^{+ a} -
\frac{(u^+\cdot q^{+})}{(u^- \cdot q^{+})} \,u^{- a}~. 
$$
The superfield $q^{+a} = f^{+a}(x,u) + \cdots = f^{ai}(x)u^+_i + \cdots $ 
is an extra compensating hypermultiplet, with the $\theta $ expansion and
reality properties entirely analogous to \p{real}, \p{thetaexp1}. Like in 
\cite{IvV}, we fully fix the local $SU(2)_c$ symmetry of \p{dTN} (which is
present in any QK sigma model action) by the gauge condition 
\beq f^{i}_a(x) =
\delta^{i}_a\, \omega(x)~. \label{gauge}  
\eeq 

The objects defined so far are necessary ingredients of the generic QK sigma
model action. The specificity of the given case is revealed in the
particular form  of ${\cal L}^{+4}$ in \p{action2}. It includes two analytic
gauge abelian  superfields $V^{++}(\zeta)$ and $W^{++}(\zeta)$ and two sets of
$SU(2)$ breaking parameters $c^{(ij)}$ and $a^{(ab)}$ satisfying 
the pseudo-reality condition  
\beq
\overline{c^{(ij)}} = \epsilon_{ik}\epsilon_{jl}c^{(kl)} \eeq             
(and the same for $a^{(ab)}$). The Lagrangian \p{action2} can be checked to
be invariant under the following two commuting gauge $U(1)$ transformations,
with the parameters $\varepsilon(\zeta)$ and $\varphi(\zeta)$:  \beqa &&\delta
Q^{+a}_A = \varepsilon \left[ \epsilon_{AB} Q^{+a}_B - \kappa^2 c^{+-}
Q^{+a}_A \right]~, \;\;\delta g^{+r} = \varepsilon \kappa^2 \left[ c^{(rn)}
g^+_n -  c^{+-} g^{+ r} \right]~, \nn  && \delta q^{+ a} = \varepsilon \kappa^2
c^{(ab)}q^+_b~, \;\; \delta W^{++} = {\cal D}^{++}\varepsilon~, \quad
(\,c^{+-} \equiv c^{(ik)}v^{+}_iu^-_k \,) \label{eps} \eeqa 
\beqa
&&\delta Q^{+a}_A = \varphi \left[ a^{(ab)}Q^+_{b\,A} - \kappa^2
(u^-\cdot g^+) Q^{+a}_A\right]~, \;\; \delta
g^{+r} = \varphi \left[ v^{+r} - \kappa^2  (u^-\cdot g^+) g^{+r} \right]~, 
\nn 
&& \delta q^{+ a} = \varphi \kappa^2 (u^-\cdot q^+)g^{+ a}~, \;\; \delta
V^{++} = {\cal D}^{++}\varphi~. 
\eeqa
This gauge freedom  will be fully fixed at the end. The only surviving 
global symmetries are two commuting $U(1)$. One of them comes from 
the Pauli-G\"ursey $SU(2)$ acting on $Q^{+a}_A$ and broken by the
constant triplet $a^{(bc)}$.  
Another $U(1)$ is the result of breaking of the
$SU(2)$ which uniformly rotates the doublet indices of harmonics and those of
$q^{+a}$ and $g^{+r}$. It does not commute with supersymmetry and forms the
diagonal subgroup  in the product of three independent  $SU(2)\,$s realized on
these quantities in the ``free'' case; this product gets broken  down to the
diagonal $SU(2)$ and further to $U(1)$ due to the presence of explicit
harmonics and constants $c^{(ik)}$ in the interaction terms in \p{action2}.
These two $U(1)$ symmetries will be isometries of the final QK
metric, the first one becoming triholomorphic in the HK limit. The fields
$D(x)$ and $V^{(ik)}_m(x)$ are inert under any isometry (modulo some rotations
in the indices $i, j$), and so are ${\cal D}^{++}$ and the $D,\; V$ part of
\p{dTN}. 

It can be shown that the action \p{dTN}, \p{action2} is a generalization of 
both the HSS quotient actions describing the QK extensions of the 
EH and Taub-NUT sigma models: putting $g^{+r} = a^{(ab)}= 0$ yields 
the EH action as it was given in \cite{bgio,IvV}, putting $Q^{+ a}_{A=2} (Q^{+
a}_{A=1}) =  c^{(ik)} = 0$ yields the Taub-NUT action \cite{IvV}. Also, fixing 
the gauge with respect to the $\lambda $ transformations by the 
condition $(u^-\cdot g^+) = 0$, varying with repect to the non-propagating 
superfield $V^{++}$ and eliminating altogether $(v^+\cdot g^+)$ by the
resulting algebraic constraint, we arrive at the form of the action which in 
the HK limit $\kappa^2 \rightarrow 0$ exactly coincides with the HSS action
describing the ``double Taub-NUT'' manifold \cite{giot,gorv}. Thus \p{dTN},
\p{action2}  is the natural QK generalization of the action of \cite{giot,gorv}
and therefore  the relevant  metric is expected to be a QK generalization of
the double Taub-NUT HK  metric.      
\vspace{0.3cm}

\noindent{\bf 3. Towards the target metric}. We are
going to profit from the opportunity  to choose a WZ gauge for $W^{++}$ and
$V^{++}$, in which harmonic differential equations for $f^{+a}(x,u),
F^{+b}_A(x,u)$ and  $g^{+ r}(x,u)$ are drastically simplified. 

In this gauge $W^{++}$ and $V^{++}$ have the following short expansion 
\beqa
&& W^{++} = i\theta^+\sigma^m\bar\theta^+ W_m(x) + (\theta^+)^2(\bar\theta^+)^2
P^{(ik)}(x)u^-_iu^-_k~, \nn
&& V^{++} = i\theta^+\sigma^m\bar\theta^+ V_m(x) + (\theta^+)^2(\bar\theta^+)^2
T^{(ik)}(x)u^-_iu^-_k \label{WZ}
\eeqa
(once again, possible terms proportional to $(\theta^+)^2$ and
$(\bar\theta^+)^2$ can be omitted). The hypermultiplet superfields have
the same expansions as in \p{thetaexp1}. At the intermediate step it is
convenient  to redefine these superfields as follows   
\beq
\left(Q^{+a}_A, \; g^{+r}\right)  = \kappa (u^-\cdot
q^+)\left(\widehat{Q}^{+a}_A, \; \widehat{g}^{+r}\right)~. \label{hat} 
\eeq
Due to the structure of the WZ-gauge superfields \p{WZ}, the highest components
in the $\theta$ expansions of the redefined HM superfields appear only in the 
kinetic part of \p{action2}. This results in the linear harmonic equations for
$f^{+a}(x,u), \widehat{F}^{+b}_A(x,u), \widehat{g}^{+r}(x,u)$: 
\beqa  
&&\partial^{++}f^{+a} = 0\; \Rightarrow \;f^{+a} = u^{+ a} \omega(x)~, \; 
\partial^{++}\widehat{F}^{+a}_a = 0\; \Rightarrow \;\widehat{F}^{+a}_A =
\widehat{F}^{ai}_A(x)u^+_i~, \nn 
&& \partial^{++}\widehat{g}^{+r}= 0\;
\Rightarrow \;\widehat{g}^{+r} = \widehat{g}^{ri}(x)u^+_i~, \label{solu}
\eeqa
where we have simultaneously fixed the gauge \p{gauge}. 

Next steps are technical and quite
similar to those explained in detail in \cite{IvV} on the  examples of the QK
extensions of the Taub-NUT and EH metrics. One substitutes the
solution \p{solu} back into the action (with the  $\theta$ and $u$ integrals
performed), varies with respect to the rest of non-propagating fields and also
substitutes the resulting relations  back into the action. At the final stages
it proves appropriate to redefine the basic fields once again
\beq     
\widehat{F}^{ai}_A = {1\over \kappa \omega}F^{ai}_A~, \quad \widehat{g}^{ri} =
{2\over \kappa \omega} g^{ri}  
\eeq
and to fully fix the residual gauge freedom of the WZ gauge for the $\varphi$ 
transformations (with the singlet gauge parameter $\varphi (x)$), so as to
gauge away the singlet part of $g^{ri}(x)$: 
\beq
g^{ri}(x) = g^{(ri)}(x)
\eeq
(the residual $SO(2)$ gauge freedom, with the parameter $\varepsilon(x)$, will
be kept for the moment). In particular, in terms of the thus defined fields we
have the following expressions for the fields  $\omega$ and $V^{(ij)}_m$
which are obtained by varying the full action \p{dTN}  with respect to $D$
and $V^{(ij)}_m$:  \beq  \kappa\, \omega = {1\over \sqrt{1 -
{\lambda\over 2} g^2- 2\lambda F^2 }}~, \quad V^{(ij)}_m = - 16
\lambda^2\omega^2 \left[ F^{a(i}_A\partial_m F^{\,j)}_{a A} + {1\over 4}
g^{r(i}\partial_m g^{j)}_r \right]~,  \eeq
where 
\beq
F^2 \equiv F^{ai}_AF_{ai A}~, \quad g^2 \equiv g^{ri}g_{ri}~, \quad \lambda
\equiv {\kappa^2\over 4}~. 
\eeq 

The final form of the sigma model Lagrangian in terms of the fields
$F^{ai}_A(x)$ and $g^{(rk)}(x)$ is as follows (we replaced
altogether ``$\partial_m$'' by ``$d$'', thus passing to the distance 
in the target QK space instead of its $x$-space pullback) 
\beq
\frac 1{{\cal D}^2}\left\{{\cal D}\left(X+Z+\frac Y4\right)+
\la\left(g^2\cdot\frac Y8+2T\right)\right\} \label{dist0}
\eeq
with
\beqa 
&& {\cal D}=1-\frac{\la}{2}\,g^2-2\la\,F\,^2,\; \;X=\frac{1}{2}\,dF_{ai\,A}\,
dF^{ai}_{A},\quad Y=\frac{1}{2}\, dg_{ij}\, dg^{ij},\nonumber \\ 
&& Z = \frac{1}{4\,\alf\,\be-\ga^2}\, 
\left\{\ga\, (J\cdot K)-\alf\, (J\cdot J)
 -\be\, (K\cdot K)\right\},\nn
&& T=F^{\,i}_{a\,B}\,dF^{aj}_{B}\left(F_{ai\,A}\,dF^{a}_{\,j\,A}
+\frac{1}{2}\,g_{ir}\,dg^r_{~j}\right)~.\label{defD}
\eeqa
Here
\beq
J=\frac 12\ a^{ab}\ F_{a\,A}^{\,i}\ dF_{bi\,A},\qq 
K=-\frac 12\,\eps_{AB}\ F^{ai}_{A}\,dF_{ai\,B}
-\frac{\la}{2}\ c_{ij}\ g^i_{~s}\ dg^{sj},
\eeq
and
\beqa 
&& \alf=\frac12\left(\frac{F\,^2}{4}-\la\,\hat{c}^2
+\frac{\la^2}{2}\,\hat{c}^2\,g^2\right),\quad
\be=\frac14\left(1+\frac{\hat{a}^2}{4}\,F\,^2 - 
\frac{\la}{2}\,g^2\right),\nn
&& \ga=\frac 14\,a^{ab}\,F^{\,i}_{a\,A}\,F_{bi\,B}\,\eps_{AB}-\la(c\cdot g)~,
\eeqa 
where 
\beq
\hat{c}^2 \equiv c^{ik}c_{ik}~, \quad \hat{a}^2 = a^{ab}a_{ab}~.
\label{squar1} 
\eeq

On top of this, there are two algebraic constraints on the
involved fields 
\beqa 
&& F^{a(i}_A\,F^{\,j)}_{a\,B}\,\eps_{AB} -\la\,g^{(li)}\,
g^{(rj)}\,c_{(lr)}+c^{(ij)}=0~,  \label{contr1} \\ 
&& g^{ij}=a^{ab}\,F^{\,i}_{a\,B}\,F^{\,j}_{b\,B}~, \label{contr2} 
\eeqa
which come out by varying the action with respect to the auxiliary fields
$P^{(ik)}(x)$ and $T^{(ik)}(x)$ in the WZ gauge \p{WZ}. Keeping in mind these 6
constraints and one residual gauge ($SO(2)$) invariance, we are
left with just four independent bosonic target coordinates as compared
with 11 such coordinates in \p{dist0}. The problem is now to explicitly solve
\p{contr1}, \p{contr2}. But before turning to this issue, let us notice that
the sought metric includes three parameters. These are the Einstein constant,
related to $\,\la,$ and two breaking parameters : the triplet $\,c^{(ij)},$
which breaks the $\,SU(2)_{\rm SUSY}\,$ to $\,U(1),$ and the triplet 
$\,a^{(ab)},$ which breaks the Pauli-G\"ursey $\,SU(2)\,$ to $\,U(1)$. The
final isometry group is therefore $\,U(1)\times U(1)$. For convenience 
we choose the following frame with respect to the broken $SU(2)$ groups 
$$c^{12}=ic,\quad c^{11}=c^{22}=0,\qq a^{12}=ia,\quad
a^{11}=a^{22}=0~,$$
with real parameters $\,a\,$ and $\,c,$ and we shift  $\dst\ \la\ \to\ 
\frac{\la}{a^2}$. Hereafter  we shall use this frame, in which, in
particular, the squares \p{squar1} become
$$
\hat{c}^2 = 2 c^2~, \quad \hat{a}^2 = 2 a^2~. 
$$
\vspace{0.3cm}   

\noindent{\bf 4. Solving the constraints}. We need to find true coordinates
to compute the metric. This step is non-trivial, due to the fact that
\p{contr1} becomes quartic  after substitution of \p{contr2}. Instead of
solving this quartic equation, it proves more fruitful to take as independent 
coordinates just the components of the triplet $g^{(ri)}$ 
$$
g^{12}= g^{21}\equiv iah,\quad \ol{h}=h,\qq
g^{11} \equiv g,\quad g^{22}=\ol{g},
$$
and one angular variable from $F^{ai}_A$. Then, relabelling the 
components of the latter fields as follows 
$$\left\{\barr{l}
\dst F^{a=1\ i=2}_{A=1}=\frac 12({\cal F}+{\cal K}),\quad 
F^{a=1\ i=1}_{A=1}=\frac 12({\cal P}+{\cal V}),\\[5mm]
\dst F^{a=1\ i=2}_{A=2}=\frac 1{2i}({\cal F}-{\cal K}),\quad 
F^{a=1\ i=1}_{A=2}=\frac 1{2i}({\cal P}-{\cal V}),\\[5mm]
\quad F^{a=2\ i=1}_{A}=-\ol{F^{a=1\ i=2}_{A}},
\quad F^{a=2\ i=2}_{A}=\ol{F^{a=1\ i=1}_{A}},\earr\right.
$$
we substitute this into \p{contr1}, \p{contr2}, and find the following general
solution (it amounts to solving a quadratic equation and we choose the
solution which is regular in the limit $g = \bar g = h =0$)
\beqa && \dst {\cal P}=-iM\,e^{i(\phi+\alf/\rho_- +\mu\rho_+)}, \qq\qq 
\dst {\cal F}=R\,e^{i(\phi+\mu\rho_-)}, \nn
&& \dst {\cal K}=iS\,e^{i(\phi-\alf/\rho_- -\mu\rho_+)}, 
\qq\qq\quad \; \dst {\cal V}=L\,e^{i(\phi-\mu\rho_-)}, 
\nn
&& \rho_{\pm}=1\pm 4\frac{\la c}{a^2} \label{sol1}
\eeqa
and
\beq
g=at\,e^{i(\alf/\rho_-+8\la c/a^2 \mu)}~. \label{sol2}
\eeq
The various functions involved are
$$\barr{ll}
\dst L=\sqrt{\frac 12(\sqrt{\De_-}+B_-)}, & \qq\qq
\dst R=\sqrt{\frac 12(\sqrt{\De_+}+B_+)},\\[4mm]
\dst M=\sqrt{\frac 12(\sqrt{\De_+}-B_+)}, & \qq\qq
\dst S=\sqrt{\frac 12(\sqrt{\De_-}-B_-)},\earr$$
with
$$\left\{\barr{cc}
\dst A_{\pm}=1\pm 2\la c\,h, & \dst\qq\qq 
B_{\pm}=c(1+\la r^2)\pm h\,A_{\mp},\\[4mm]
\dst \Delta_{\pm}=B^2_{\pm}+t^2\,A_{\mp}^2, & \dst\qq r^2=h^2+t^2~, \;\;
g\bar g = a^2 t^2. \earr\right.$$
The true coordinates are $\,(\phi,\,\alf,\,h,\,t)$. 
An extra angle $\mu$ parametrizes the local $SO(2)$ transformations (they
act as shifts of $\mu$ by the parameter $\varepsilon(x)$). In view of the gauge
invariance  of \p{dist0}, the final form of the metric should not depend on
$\mu $ and we can choose the latter at will. For instance, we can change the
precise dependence of phases in \p{sol1}, \p{sol2} on $\phi$ and $\alpha$. In
what follows we shall stick just to the above parametrization.  
\vspace{0.3cm} 

\noindent{\bf 5. The resulting metric}. To get the full metric is fairly 
involved and Mathematica was intensively used ! The final result is
\beq\label{met1}
g=\frac 1{4{\cal D}^2}\ \frac{\cal P}{\cal A}
\left(d\phi+\frac{\cal Q}{4{\cal P}}d\alf\right)^2+
\frac{\cal A}{{\cal D}^2}
\left(dh^2+dt^2+\frac{t^2}{\cal P}(1+\la r^2)^2\,d\alf^2\right).\eeq
It depends on 4 functions
$${\cal D},\qq {\cal A},\qq {\cal P},\qq {\cal Q},$$
where
$$\barr{ll}\dst {\cal A}=\frac{a^2}{4}+\frac 18(1-4\la c^2)(1-\la\, r^2)
\left(\frac 1{\sqrt{\De_+}}+\frac 1{\sqrt{\De_-}}\right)\\[6mm]
 \dst \hspace{4cm}
-\la\,c\, h\left(\frac 1{\sqrt{\De_+}}-\frac 1{\sqrt{\De_-}}\right) 
+\frac{\la c^2}{a^2}\,
\frac{4\la\, t^2-(1+\la\,r^2)^2}{\sqrt{\De_+}\sqrt{\De_-}},\earr$$
$$\barr{l}
\dst {\cal P}=(1+\la\,r^2)^2\left(1-
\frac{2\la\,c}{a^2}
\left(\frac{h+c(1-\la\,r^2)}{\sqrt{\De_+}}-
\frac{h-c(1-\la\,r^2)}{\sqrt{\De_-}}\right)\right)^2\\[6mm]
\dst\hspace{4cm}+\frac{4\la^2\,c^2\,t^2}{a^4}
\left(\frac{1-\la\,r^2-4\la\,c\,h}{\sqrt{\De_+}}-
\frac{1-\la\,r^2+4\la\,c\,h}{\sqrt{\De_-}}\right)^2,\earr$$
$$\barr{l} 
\dst {\cal Q}=-(1+\la\,r^2)^2\left(\frac{h+c(1-\la\,r^2)}{\sqrt{\De_+}}+
\frac{h-c(1-\la\,r^2)}{\sqrt{\De_-}}\right) \\[6mm] \dst\hspace{5cm}
+4\la\,c\,t^2\left(\frac{1-\la\,r^2-4\la c\,h}{\sqrt{\De_+}}
-\frac{1-\la\,r^2+4\la c\,h}{\sqrt{\De_-}}\right),\earr$$
and $${\cal D}=1-\la\,r^2-2\frac{\la}{a^2}\left(\sqrt{\De_+}+
\sqrt{\De_-}\right).$$ 

To simplify matter we first rescale  $\,c\to
c/2$. The relations $$ \De_{\pm}=(1+\la\,c^2)\,t^2+(h \pm c/2(1 - \la\,r^2))^2
$$ suggest the following change of coordinates \beq
T=\frac{2t}{1-\la\,r^2},\qq\quad H=\frac{2h}{1-\la\,r^2}, \qq
\rho=\sqrt{T^2+H^2}, \label{1chan} \eeq  which has the virtue of reducing
the quartic non-linearities according to
$$\Delta_{\pm}=\frac{(1-\la\,r^2)^2}{4}\,\de_{\pm},\qq\quad
\de_{\pm}=(1+\la\,c^2)\,T^2+(H\pm\,c)^2.$$ 
Further, to get rid of the square
roots we use spheroidal coordinates 
 $\,(s,x)\,$ defined by
$$\sqrt{1+\la\,c^2}\ T=\sqrt{(s^2-c^2)(1-x^2)},\qq\quad H=\,s\,x,
\qq\quad s\geq c,\qq x\in\ [-1,+1].$$
For convenience reasons we scale the angles $\,\phi\,$ and 
$\,\alf\,$ according to
$$\frac{\phi}{\sqrt{1+\la\,c^2}}\quad\Rightarrow\quad\phi,\qq\quad
\frac{\alf}{\sqrt{1+\la\,c^2}}\quad\Rightarrow\quad\alf,$$
and to have a smooth limit for $\,a\to 0\,$ we come back to the original 
$\lambda$, $\ \la\,\to\,\la\,a^2$.

Putting these changes together, we get the final form of the metric
\beq\label{f3}
\barr{l}
\dst (4l^2)\,g=(1+\la\,a^2\,s^2)\ \frac PA
\left(d\phi+\frac{Q}{4P}\,d\alf\right)^2
+\frac AP\ (s^2-c^2)(1-x^2)(1+\la\,a^2\,c^2\,x^2)\,(d\alf)^2\\[5mm]
\dst\hspace{4cm}
+A\left(\frac{ds^2}{(s^2-c^2)(1+\la\,a^2\,s^2)}
+\frac{dx^2}{(1-x^2)(1+\la\,a^2\,c^2\,x^2)}\right),\earr\eeq
with
$$\left\{\barr{l}
\dst l=1-2\la\,s,\qq Q=-2(1+\la\,a^2\,c^2)(s^2-c^2)\,x,\\[5mm]
\dst 4\,A=(2+a^2\,s)(s-2\la\,c^2)-a^2\,c^2\,l^2\,x^2,\\[5mm]
\dst P=c^2(1-x^2)(1+\la\,a^2\,c^2\,x^2)\,l^2+
(s^2-c^2)\left[1+\la\,a^2\,c^2\,x^2-4\la^2\,c^2(1-x^2)\right].
\earr\right.$$
 The isometry group $\,U(1)\times U(1)\,$ acts as translations
of 
 $\,\phi\,$ and $\,\alf.$
\vspace{0.3cm}

\noindent{\bf 6. Geometric structure of the metric}.
We know that this metric is QK by construction, but in view of 
the many steps involved, it is a good self-consistency check to verify 
that it is Einstein with self-dual Weyl tensor. The details will be
presented in \cite{civ}, let us describe the main result.
We take for the vierbein 
$$e_0=a(s,x)\ \left(d\phi+\frac{Q}{4P}\,d\alf\right),\qq
e_3=b(s,x)\ d\alf,\qq e_1=\chi\ ds,\qq e_2=\nu\ dx,$$
with 
$$\left\{\barr{ll}
\dst a(s,x)=\frac 1{2l}\ \sqrt{1+\la s^2}\ \sqrt{\frac PA},\qq & 
\dst\chi=\frac 1{2l}\ \sqrt{\frac A{\cal C}},
\qq {\cal C}=(s^2-c^2)(1+\la\,s^2),\\[5mm]
\dst b(s,x)=\frac 1{2l}\ \sqrt{(s^2-c^2){\cal B}}\ \sqrt{\frac AP},\qq &
\dst \nu=\frac 1{2l}\ \sqrt{\frac A{\cal B}},
\qq {\cal B}=(1-x^2)(1+\la\,c^2\,x^2).\earr\right.$$
The spin connection being defined as usual by
$$de_a+\om_{ab}\wedge e_b=0,\qq a,b=0,1,2,3,$$
one has to compute the anti-self-dual spin connection and curvature
$$\om^-_i=\om_{0i}-\frac 12\,\eps_{ijk}\,\om_{jk},\quad R^-_i\equiv R_{0i}-\frac 12\,\eps_{ijk}\,R_{jk}=d\om^-_i+\eps_{ijk}\ \om^-_j\wedge\om^-_k,\quad i,j,k=1,2,3.$$
One gets the crucial relation
\beq\label{Einstein}
R^-_i=-16\la\left(e_0\wedge e_i-\frac 12\,\eps_{ijk}\ e_j\wedge e_k\right),\eeq
which shows at the same time that the metric is Einstein, with
$${\rm Ric}=\Lambda\ g,\qq\qq \frac{\Lambda}{3}=-16\la = - 4 \kappa^2,$$
and that the Weyl tensor is self-dual, i.e. $\ W^-_i=0.$

Let us now consider a few limiting cases. 
\vspace{0.3cm}

\noindent{\it The quaternionic Taub-NUT limit}.
Let us show that in the limit $\,c\to 0\,$ we get the 
quaternionic Taub-NUT. We first write the metric (\ref{met1}) in the form
$$g(c\to 0)=\frac 1{4{\cal D}^2}\left\{
\frac{(1+\la\,r^2)^2}{{\cal A}_0}\left(d\psi+\frac hr\,d\alf\right)^2
+{\cal A}_0\,\ga_0\right\},$$
with
$$\left\{\barr{l}
\dst \psi=-2\,\phi,\qq{{\cal A}_0}=a^2+\frac 1r-\la r,\qq
{\cal D}=1-\la\,r^2-4\frac{\la r}{a^2},\\[5mm]
\ga_0=dh^2+dt^2+t^2\,d\alf^2,\qq\qq r^2=h^2+t^2.\earr\right.$$
Switching to the spherical coordinates $\,r,\,\tht,\,\alf\,$ for which
$$t=r\sin\tht,\qq h=r\cos\tht$$
allows one to get the final form
\beq\label{q1}
g(c\to 0)=\frac{(1+\la\,r^2)^2}{4{\cal D}^2}\left\{
\frac 1{{\cal A}_0}\,\si_3^2
+\frac{{\cal A}_0}{(1+\la\,r^2)^2}(dr^2+r^2(\si_1^2+\si_2^2))\right\},\eeq
with
$$\si_3=d\psi+\cos\tht\,d\alf,\qq\quad 
\si_1^2+\si_2^2=d\tht^2+\sin^2\tht\,d\alf^2.$$

The derivation of the quaternionic Taub-NUT metric from harmonic superspace was 
given in \cite{iv1}. It contains 2 parameters $\,\tilde{\la},\  R,\,$ and in the 
limit $\,R\to 0\,$ it reduces to Taub-NUT. One can see that, upon
the identifications $$s=r,\qq a^2=4\tilde{\la}^2,\qq \la=-R\,\tilde{\la}^2,$$
the metric $\,2g(c\to 0)\,$ is nothing but the quaternionic Taub-NUT.
\vspace{0.3cm}

\noindent{\it The quaternionic Eguchi-Hanson limit}.
This metric was derived using harmonic superspace in \cite{IvV}, and can 
be written as
\beq\label{eh1}
4C^2\,g=\frac{(\tilde s^2-\tilde c^2)}{\tilde sB}\,\tilde{\si}_3^2+\tilde sB\left(\frac{d\tilde s^2}{\tilde s^2-\tilde c^2}+
\tilde{\si}_1^2+\tilde{\si}_2^2\right),\eeq
where
$$ \tilde sB=\tilde s-\kappa^2\,\tilde c^2,\quad C=1-\kappa^2\,\tilde s,\qq \tilde{\si}_3=d\phi+
\cos\tht\,d\psi,\quad \tilde{\si}_1^2+\tilde{\si}_2^2=d\tht^2+
\sin^2\tht\,d\psi^2.$$
The writing (\ref{eh1}) is adapted to the Killing $\,\partial_{\phi};$ if we 
switch to the Killing $\,\partial_{\psi}\,$ we can write the metric as
\beq 
\frac D{4\tilde sBC^2}(d\psi+\EuFrak{B})^2+\frac{\tilde
sB}{4C^2}\left(\frac{d\tilde s^2}{\tilde s^2-\tilde c^2} +d\tht^2+\frac{\tilde
s^2-\tilde c^2}{D}\,\sin^2\tht\,d\phi^2\right), \label{eh11} 
\eeq 
with
$$D=(\tilde s^2-\tilde c^2)\cos^2\tht+(\tilde sB)^2\,\sin^2\tht,\qq\EuFrak{B}=\frac{\tilde s^2-\tilde c^2}{D}\,
\cos\tht\,d\phi.$$

If we now take, in the metric (\ref{f3}), the limit $\,a\to 0\,$ it 
becomes proportional to the metric (\ref{eh11}) upon the following
identifications $$s\; = \; 2\tilde s,
\qq c\; =\; 2\tilde c,
\qq \la\; =\; \frac{\kappa^2}{4},
\qq \phi\; \rightarrow \; \frac{\psi}{2},
\qq \alf\;\rightarrow \; -\phi,
\qq x \; \rightarrow\; \cos\tht.
$$
\vspace{0.3cm}

\noindent{\it The hyper-K\" ahler limit}.
Relation (\ref{Einstein}) makes it clear that in the limit $\,\la\to
0\,$ we recover a Riemann self-dual geometry, which is therefore  hyper-K\"
ahler. At the level of the metric, it is most convenient to discuss it using
the co-ordinates (\ref{1chan}). Indeed, we obtain the multicentre structure
\cite{ksd,gh,Hi}  $$\frac 14\left[\frac 1V\,(d\phi+{\bf
A})^2+V\,\ga_0\right],$$  with the flat 3-metric
$$\ga_0=dH^2+dT^2+T^2\,d\alf^2~.$$
The potential $\,V\,$ and the connection $\,{\bf A}\,$ are, respectively,
\beqa 
V &=& \frac{1}{4}\left\{a^2+\frac 1{\sqrt{\de_+}}+\frac
1{\sqrt{\de_-}}\right\},\qq {\bf A}=-\frac 14\left\{\frac{H+c}{\sqrt{\de_+}}+
\frac{H-c}{\sqrt{\de_-}}\right\}\,d\alf~,\label{hk} \\
\de_{\pm}&=& T^2+(H\pm c)^2~. \nonumber \eeqa
The potential shows two centres
and $\,V(\nf)=a^2/4.$ An easy computation  gives 
\newcommand{\dec}{\mathop{\smash{\star}}}
$$dV=-\dec_{\ga_0}\, d{\bf A}~,$$
which is the fundamental relation of the multicentre metrics. For 
$\,a\neq 0\,$ we have the double Taub-NUT metric, while for $\,a=0\,$ 
we are back to the EH metric.
\vspace{0.3cm}

\noindent{\it Comparison with other known QK metrics}. The QK metric
considered here is Einstein with self-dual Weyl tensor. From a general result
due to Przanowski \cite{Pr}  and Tod \cite{To}, this class of metrics is  
conformally related to a subclass of K\"ahler scalar-flat ones. From a result 
of Flaherty \cite{Fl}, any  K\"ahler scalar-flat metric is a solution of the 
coupled Einstein-Maxwell equations, with the restriction that the Weyl
tensor be self-dual. The explicit solutions of the coupled
Einstein-Maxwell equations known so far fall in two classes: the
Perj\'es-Israel-Wilson metrics \cite{iw,Pe} and  the 
Plebanski-Demianski \cite{pd} metrics. In general they are not Weyl-self-dual.

For the first class we have checked (details will be given in \cite{civ}),
that the Weyl-self-dual metrics are conformal to the
multicentre metrics. For the metrics in the second class, imposing  Weyl
self-duality indeed gives rise to a QK metric. In the HK limit, with the same
coordinates as in (\ref{hk}), we have found its potential to be $$V=\frac
1{\sqrt{\de_+}}+\frac m{\sqrt{\de_-}}~.$$ 
For $\,m=0\,$ we recover flat space while for $\,m\neq 1\,$ it describes a 
deformation of Eguchi-Hanson with two unequal masses. Thus our metric is 
also outside the Plebanski-Demianski ansatz, since their HK limits
are  different. We conclude  that it supplies a novel explicit example of 
the Einstein metrics with the self-dual Weyl tensor and, simultaneously, of 
the solution of the coupled Einstein-Maxwell system.
\vspace{0.3cm}

\noindent{\bf Acknowledgments}. G.V. thanks Gary Gibbons for useful
discussions and for bringing to his attention the references
\cite{iw,pd}. Work of E.I. was partially supported by grants RFBR-CNRS
98-02-22034,  RFBR-DFG-99-02-04022, RFBR 99-02-18417 and NATO grant PST. CLG
974874. He thanks the Directorate of LPTHE for the hospitality extended to him
within the Project PAST-RI 99/01.

\end{document}